\documentstyle[twocolumn,aps,prb]{revtex}
\begin{document}
\draft
\title{Electron scattering on circular symmetric magnetic profiles in a two-dimensional electron gas}
\author{J. Reijniers, F. M. Peeters \cite{peeters} and A. Matulis \cite{matulis}}
\address{Departement Natuurkunde, Universiteit Antwerpen (UIA), \\
Universiteitsplein 1, B-2610 Antwerpen, Belgium}

\date{\today}
\maketitle
\begin{abstract}
The quasi-bound and scattered states in a 2DEG subjected to a
circular symmetric steplike magnetic profile with zero average
magnetic field are studied. We calculate the effect of a random
distribution of such identical profiles on the transport
properties of a 2DEG. We show that a nonzero Hall resistance can
be obtained, although $\langle B_{z} \rangle=0$, and that in some
cases it can even change sign as function of the Fermi energy or
the magnetic field strength. The Hall and magnetoresistance show
pronounced resonances apart from the Landau states of the inner
core, corresponding to the so-called quasi-bound snake orbit
states.
\end{abstract}
\pacs{}

\section{Introduction}

The response of a two-dimensional electron gas (2DEG) to a
spatially inhomogeneous magnetic field has been the subject of
considerable interest in recent years.\cite{peeters99} At very low
temperatures and in very pure samples, these inhomogeneities in
the magnetic field can act as scattering centers for the 2DEG,
perturbing the ballistic electron motion and hence altering the
transport properties of the 2DEG.

These inhomogeneous magnetic fields can be realized by growing a
type-II superconducting film on top of a heterojunction,
containing a 2DEG.\cite{geim} If a background magnetic field is
applied, vortices will penetrate the 2DEG, where they form
scattering centers. If the applied magnetic field is low, the
vortices will be distributed randomly, due to the inhomogeneities
in the superconducting film. Brey \emph{et al.}\cite{brey93} and
Nielsen \emph{et al.}\cite{nielsen95} studied scattering on these
vortices if distributed randomly, and if distributed on a periodic
array.

In an alternative approach superconducting particles are
deposited above a 2DEG. Due to the Meissner effect, magnetic flux
will be expelled from the particles, which again results in a low
magnetic field region underneath each of the superconducting
particles. This was realized by Smith \emph{et
al.},\cite{smith94} who grew lead grains on top of a
heterojunction.

A logic next step would be to deposit ferromagnetic clusters as
inhomogenous magnetic field creators in the 2DEG. This was
realized by Ye \emph{et al.},\cite{ye97} who grew Dy-micromagnets
on top of a GaAs/AlGaAs heterostructure and recently Dubonos
\emph{et al.}\cite{dubonos99} studied scattering of electrons on
the stray field of a single Dy-magnet. This problem is
essentially different from the earlier problems, because now, the
average magnetic field strength is zero, $\langle B_z \rangle =0$.

Preliminary results on this system were already presented in
Refs.~\onlinecite{peeters96} and \onlinecite{reijniers00}, where
scattering on the stray fields of infinitesimaly flat magnetic
disks with perpendicular magnetization was studied. In this paper
we extend and generalize these earlier results and study
cylindrical symmetric steplike profiles, with average magnetic
field zero. This simplification enables us to classify the arising
phenomena and understand their underlying physics. We will show
that such a system can give rise to a nonzero Hall resistance,
even though $\langle B_z \rangle = 0$. Moreover, such a system can
host quasi-bound states, similar to the ones studied by Kim
\emph{et al.},\cite{kim01} who investigated theoretically the
electron states of a circular symmetric magnetic field profile,
consisting of two regions with different magnetic field strengths
inside and outside a radius $R$. In contrast to the bound states
found by Kim \emph{et al.}, here the electron states are
quasi-bound, because we consider magnetic profiles which are
finite in extent.

We will model the magnetic field by
\begin{eqnarray}
B(r<R_a)&=&B_a \nonumber \\
B(R_a<r<R_b)&=&B_b \\
B(r>R_b)&=&0\nonumber,
\end{eqnarray}
with $B_b=-B_a/[(R_b/R_a)^2-1]$ such that the condition $\langle
B_z\rangle =0$ is satisfied. This magnetic field profile models
the one of a perpendicularly magnetized ferromagnetic disk as
felt by a 2DEG underneath the disk. As an example we plot in
Fig.~\ref{fig:magnetic.eps} the magnetic field profile (solid
curve) resulting from a ferromagnet with radius $R_a$ and
thickness $d/R_a=1$, grown a distance $h/R_a=0.1$ above a 2DEG,
as shown in the inset. The dotted curve represents the magnetic
field according to our model, where we have chosen
$R_b/R_a\approx 2.8$ to account for the shape of the profile. The
cut-off at $R_b$ is made in order to simplify the calculations.
We believe that in doing so, the physics is not altered.

The parameters ($R_a$, $R_b$, $B_a$ and $B_b$) depend on the
specific properties of the ferromagnetic material (as extent,
thickness, magnetization, distance to the 2DEG). The resulting
profile can also be affected by including superconducting strips,
which expel the flux lines due to the Meissner effect and
consequently can rearrange/guide the magnetic field lines.

The paper is organized as follows. In Sec.~II we consider
scattering on a single magnetic profile. First, we solve the
problem classically in order to get a reference frame which
describes the large energy limit. Then we concentrate on the
quantum mechanical behaviour, and study the differences which
arise, as e.g. the existence of quasi-bound states. In Sec.~III,
we calculate the response of the 2DEG to a random, homogeneous
distribution of these (identical) profiles over the sample. The
approach is along the lines presented by Nielsen and Hedeg\aa rd
in Ref.~\onlinecite{nielsen95}, which was based on the Boltzmann
transport equation. Again, we consider the scattering both
classically and quantum mechanically, and discuss the arising
differences. In Sec.~IV we summarize our conclusions and discuss
briefly the possibility of reproducing our results experimentally.

\section{Scattering on a single magnetic profile}
\subsection{Classical scattering}
Classically, the scattering on a magnetic field profile is
determined by the solution of Newton's equation of motion, where
the force is given by the Lorentz expression ${\mathbf
F}=e{\mathbf v}\times{\mathbf B}$ for a particle with charge $e$.
Outside the profile, no magnetic field is present, and
consequently the path is just a straight line. Inside the profile,
the electron describes an arc of a circle, of which the radius
(and the direction in which it is drawn, i.e., clock- or
counterclockwise) depends on its position in the profile, since
the profile considered here consists of two regions with different
magnetic field strength (and sign). The respective cyclotron
radii are given by $l_c=v/\omega_c$, where $v$ is the velocity of
the electron and $\omega_c=eB/mc$ is the cyclotron frequency in
the local magnetic field, which is $B=B_a$ in the inner core and
$B=B_b$ in the outer region.

The geometry of the scattering process is determined by the
following dimensionless parameters: (a) $R_b/R_a$, i.e., the ratio
of the radii of the inner and the outer circle of the magnetic
field profile, and (b) $l_a/R_a=(m/e)(v/B_a)$, which is the ratio
of the cyclotron radius in the inner core, to the radius of this
center part. It is clear from geometrical considerations, that it
is impossible for a particle which was initially outside the
magnetic profile to become trapped inside the magnetic profile.

We calculated the differential cross section $d\sigma/d\phi$
numerically, from the different classical trajectories. In
Fig.~\ref{fig:trajectories.eps}, we show examples of the
classical trajectories (on the right) and their resulting cross
sections (bold curves, on the left) for the $R_b/R_a =1.5$
configuration for different $l_a/R_a$.

In the limit of $l_a/R_a\rightarrow 0$, the cyclotron radius is
very small compared to $R_b$, and therefore the electron scatters
on the magnetic profile as if it were bouncing off a hard wall.
As a consequence the differential cross section would be
symmetric in $\phi$. From Fig.~\ref{fig:trajectories.eps}, we see
that if $l_a/R_a$ increases, the differential cross section
changes drastically, and looses its symmetry due to the time
reversal breaking magnetic field. We notice that $d\sigma/d\phi$
is sensitive to $l_a/R_a$: for increasing $l_a/R_a$ it is more
centered around 0, but its structure also changes significantly,
which can be understood by inspection of the different classical
electron trajectories shown on the right.

Suppose we consider the classical trajectories from left to
right. In the case $l_a/R_a=0.5$, the electrons are deflected to
the left ($\phi<0$), giving rise to the orbits indicated by (1),
and contributing to the differential cross section indicated by
(1). Shifting the initial position of the electron further to the
right, at a certain point the electron does not only feel the
outer region, but is able to penetrate into the inner core, and
consequently the electron is abruptly swept to the other side.
The latter causes the abrupt decrease of $d\sigma / d\phi$ at
$\phi/\pi=-0.75$. These trajectories contribute to part (2) in
the differential cross section, and have mainly $\phi>0$.

The same reasoning can be used to understand the differential
cross section for $l_a/R_a=1$, only now, $d\sigma/d\varphi$ is
more centered around zero due to the higher velocity of the
electron (with respect to the magnetic field), i.e., the electron
gets less deflected.

For larger velocities, e.g. for $l_a/R_a=2$ as in
Fig.~\ref{fig:trajectories.eps}(c), the previous picture has to be
extended with a new type of trajectories: those on the right (3)
which again only probe the outer region, since the magnetic field
in the outer region is not strong enough to deflect the electron
into the inner part. These trajectories give rise to an additional
peak (3) in the differential cross section.

\subsection{Quantum mechanical scattering}
\subsubsection{The Schr\"{o}dinger equation}
We have to solve the following Schr\"{o}dinger equation
\begin{equation} (H-E)\Psi(r,\varphi)=0,
\end{equation}
where $E=\hbar^2 k^2/2m$ is the energy of the scattering wave.
Because of cylindrical symmetry, we work with polar coordinates
${\mathbf r}=\left( r,\varphi\right)$. We can make this equation
dimensionless by rescaling the problem in the following way:
length $R_0=R_a$, energy $E_0=\hbar^2/(mR_a^2)$, time
$t_0=mR_a^2/\hbar$ and magnetic field $B_0=c\hbar/eR_a^2$. We can
write the scattering wavefunction as consisting of components,
separated into an angular and a radial part
\begin{equation}
\Psi(r,\varphi)=\sum_{m=-\infty}^{\infty}
R_{km}(r)\Phi_m(\varphi),
\end{equation}
for which the angular part equals
\begin{equation}
\Phi_{m}(\varphi)=\frac{1}{\sqrt{2\pi}}e^{im\varphi},
\end{equation}
since the problem is cylindrical symmetric. The Schr\"{o}dinger
equation is then reduced to only one dimension
\begin{equation}
 \left[
-\frac{1}{2r}\frac{d}{dr}r\frac{d}{dr}+V_{m}(r)-E\right]
R_{km}(r)=0, \label{potvolkoffie}
\end{equation}
with
\begin{equation}
V_{m}(r)= \frac{1}{2}\left[ A_\phi(r)+\frac{m}{r} \right]^2,
\end{equation}
the effective potential and $A_\phi(r)=(1/r)\int_0^rdr'r'B(r')$
the angular component of the vector potential. Notice that the
angular quantum number satisfies $-\infty<m<\infty$, in contrast
to when the scatterers are non magnetic, in which we have $0 \le
m<\infty$, due to symmetry, i.e. $m$ and $-m$ result in the same
scattered wave.

We know that the scattering process is fully determined, if we
know $\delta_{m}$ for every $m$. In order to calculate these
phase shifts we have to solve Eq.~(\ref{potvolkoffie}) for every
$m$ in the presence of our magnetic profile, and compare the
scattered wave with the unperturbed one. The solution for
$r<R_{a}$ is
\begin{eqnarray}
R_{km}(r)&=&
r^{|m|}e^{-\frac{1}{2}B_{a}r^2}[c_{m,1}M(\alpha,\beta,\gamma)]\hspace{1.2cm}
\end{eqnarray} and for $R_{a}<r<R_{b}$
\begin{eqnarray}
R_{km}(r)&=&r^{|m|}e^{-\frac{1}{2}B_{a}r^2}\times\\
&&\hspace{0.5cm}[c_{m,2}M(\alpha',\beta,\gamma')+c_{m,3}U(\alpha',\beta,\gamma')],
\nonumber
\end{eqnarray}
%
%
where $M(a,b,c)$ and $U(a,b,c)$ are confluent hypergeometric
functions,\cite{abramo} $c_{m,i}$ are normalization constants,
$\alpha=[|m|+1-m-k^2/(2B_{a})]/2$, $\beta=|m|+1$ and
$\gamma=B_{a} r^2$. $\alpha'$ and $\gamma'$ are the same as
$\alpha$ and $\beta$ but with $B_a\rightarrow B_b$. The constants
$c_{m,2}$ and $c_{m,3}$ have to be determined such that the
wavefunction is continuous at $r=R_{a}$. An alternative approach
is to calculate the wavefunction numerically up to $r=R_b$.

For $r>R_b$, Eq.~(\ref{potvolkoffie}) reduces to
\begin{equation}
\left[ -\frac{1}{2r}\frac{d}{dr}r\frac{d}{dr}+\frac{m^{2}}{2r^{2}}%
-\frac{k^{2}}{2}\right] R_{km}(r)=0, \label{schrodingeroneindigba}%
\end{equation}
which is the differential equation for the Bessel function of the
first kind
\begin{equation}
R_{km}(r)=a_{m}J_{m}(kr)+b_{m}Y_{m}\left( kr\right). \label{radiaalb}%
\end{equation}
Therefore, the resulting phase shifts can be calculated at this
point, $r=R_b$, and are derived from the condition that the
logarithmic derivative of the radial wavefunction must be
continuous at this boundary
\begin{equation}
\left. \frac{1}{R_{km}^<}\frac{dR_{km}^<}{dr}\right|_{r=R_b}=
\left. \frac{1}{R_{km}^>}\frac{dR_{km}^>}{dr}\right|_{r=R_b}
\equiv \xi_{km},
\end{equation}
which results in
\begin{equation}
\left.
\frac{1}{R_{km}^<}\frac{dR_{km}^<}{dr}\right|_{r=R_b}=\frac{j_m(kR_b)-y_m(kR_b)\tan\delta_m}{J_{m}(kR_b)-Y_{m}(kR_b)\tan\delta_m},
\end{equation}
where we have introduced the abbreviations
$z_m(x)=(x/2)[Z_{m-1}(x)-Z_{m+1}(x)]$ with $(z,Z)$ = $(j,J)$ or
$(y,Y)$. It is now easy to solve for $\delta_m$
\begin{equation}
\tan\delta_m=\frac{j_m-\xi_{km} J_m(kR_b)}{y_m-\xi_{km}Y_m(kR_b)}.
\end{equation}

\subsubsection{Resonances}
In contrast to the classical problem the ratio $l_a/R_a$ no
longer determines the scattering problem completely. We need to
know the exact energy and the magnetic field strength, and
therefore $l_a/R_a$ should be extended with the exact $E$ or
$B_a$. We have chosen to fix the magnetic field strength
$B_a/B_0=20$, and to plot all curves as function of
$k/B_a(R_aB_0)=l_a/R_a$, as was done in the classical case. The
larger the $k$-value ($\sim l_a/R_a$ for fixed $B_a$), the more
classical the system is, and the more the average of it converges
to our previously obtained classical result (see thin solid
curves in the right part of Fig.~\ref{fig:trajectories.eps}). But
for lower $k$-values, i.e., when the wavelength of the scattering
wave is comparable to the dimensions of the scatterer $1/k
\approx R_b$, quantum mechanics becomes important and results in
features which cannot be understood classically, as e.g. the
existence of resonances.

In order to determine at which energies these resonances occur,
one should inspect the phase shift $\delta_m$ as function of the
energy or the corresponding $k$-vector. When a jump of $\pi$
occurs, there is a resonance for that particular $k$-value for
given $m$. The lifetime of that quasi-bound state depends on the
energy-interval over which this jump occurs, or on the peak width
of the partial cross section $\sigma_m$ as function of the energy.

In the following, we show the results for the case $B_a/B_0=20$
and $R_b/R_a=1.5$. In Fig.~\ref{fig:effpot.eps} we plot the
effective potential for four different $m$-values for this case.
In Fig.~\ref{fig:phase.eps} we plot the phase shift as function of
$k/B_{a}$ for $-10\leq m \leq 10$.

We notice that for $m \ge -2$ well defined quasi-bound states are
formed at the Landau levels of the inner core of the magnetic
field profile, i.e. at $k=\sqrt{B_{a}(2n+1)}=4.471$, 7.745, 10,
$\cdots$, or in the units of Fig.~\ref{fig:phase.eps}: $k/B_{a}=
0.224$, 0.387, 0.5, $\cdots$. Landau states in the outer region
($B_b/B_0=16$) are not possible since the total extent of the
lowest Landau state ($=2l_b=2R_a/\sqrt{20}=0.5R_a$) does not fit
into the outer region ($R_b-R_a=0.5R_a$).

Nevertheless, for $m<-4$ there are also resonances which have an
energy lower than the first Landau level of the inner core. They
correspond to quasi-bound snake orbit states, which travel around
the profile, propagating from the $B_a$-region into the
$B_b$-region and vice versa. As an example we plot the effective
potential and the corresponding radial wavefunction $R_m$ for
$m=-8$ in Fig.~\ref{fig:boundstate.eps}. Because the two wells,
corresponding with the magnetic confinement in the different
magnetic fields of the inner and outer regions, are joined
together, they form one well which is broader --and consequently
has a energy lower-- than each of the separate wells. A similar
effect we encountered in a previous paper for the case of
electron traveling along a 1D magnetic interface.\cite{reijniers}
The electrons propagate classically as schematically depicted in
the inset of Fig.~\ref{fig:boundstate.eps}.

For higher energies we also notice resonances for negative $m$,
e.g. for $m=-21$ as shown in Fig.~\ref{fig:boundstate.eps}. They
too correspond to snake orbits, but because they have a larger
energy, they have to move closer to the interface, in order not
to escape the magnetic field profile, since their cyclotron
radius is larger. These type of states become extinct when the
cyclotron radius in the outer part exceeds the radius of the
outer part $R_b$, i.e. for $kl_{a}\approx R_b/R_a=1.5$. We have
checked this and these resonances indeed disappear.

\subsubsection{The differential cross section}
In two dimensions, the differential cross section is given by
\begin{equation}
\frac{d \sigma}{d\phi}=\frac{2}{\pi k}\left|
\sum_{m=-\infty}^{\infty}e^{im\varphi}e^{-i\delta_m}\sin\delta_m
\right|^2. \label{eq:shittie}
\end{equation}
We plot this together with its classical counterpart in
Figs.~\ref{fig:trajectories.eps} (a-c). We notice that many
oscillations are present, due to interference effects. The number
of these oscillations depends on the energy: the larger the
energy the more oscillations. In the high energy limit, the
quantum mechanical result will ultimately, on the average,
converge to the classical one, except for the peak at $\phi=0$.
Its occurrence is a purely quantum mechanical effect, and is due
to the fact that the electron which would classically pass by
--and hence does not interact with-- the scatterer, quantum
mechanically has a finite overlap with the scatterer, and
consequently contributes --although very little-- to the cross
section. Because the interaction is very little, it gets only
scattered over a very small angle, and thus adds to the $\phi=0$
peak.

\subsubsection{The total cross section}
In Fig.~\ref{fig:tot_cross.eps} we plot the total cross section
$\sigma=\int d\phi d\sigma/d\phi$ as function of $k/B_{a}$.
Classically, it is equal to the total diameter of the magnetic
inhomogeneity $\sigma=2R_{b}$ (dashed line in
Fig.~\ref{fig:tot_cross.eps}).

From Fig.~\ref{fig:tot_cross.eps} we notice that the quantum
mechanical cross section (solid curve) is larger than the
classical result. For large energies the total cross section is
twice as large, for small energies the cross section is four
times as large, as is the case for scattering on a spherical hard
wall. We also notice the resonances mentioned before, present as
small peaks, which can be attributed to a particular $m$-value.
As an example, we indicated the $m=-8$ and $m=-21$ resonances,
corresponding to the quasi-bound states of
Fig.~\ref{fig:boundstate.eps}.

\section{Scattering on multiple profiles} With the knowledge of the classical and
quantum mechanical differential cross sections, it is now
possible to calculate the Hall and magnetoresistance in a 2DEG
subjected to a randomly distributed array of such identical
profiles. We will make the assumption that the dimensions of the
magnetic disks are small compared to the distance between the
disks, so we do not include interference effects between
different scattering events. Moreover, we neglect impurity
scattering.

We solve the (classical) Boltzmann transport equation, linearized
in the electric field, and follow the derivation as described in
the paper of Nielsen and Hedeg\aa rd.\cite{nielsen95} Finally, we
arrive at
\begin{mathletters}
\begin{eqnarray}
\rho_{xx}&=&\frac{1}{(2\pi)^2} \frac{n_0}{n_e} \frac{\hbar}{e^2}
\int_{-\pi}^{\pi} d\phi (1-\cos \phi )w(k,\phi ),\label{eq:jonassie}\\
\rho_{xy}&=&\frac{1}{(2\pi)^2} \frac{n_0}{n_e} \frac{\hbar}{e^2}
\int_{-\pi}^{\pi} d\phi \sin\phi w(k,\phi),\label{eq:jonassie2}
\end{eqnarray}
\end{mathletters}
where $n_e$ is the electron concentration, $n_0$ is the
concentration of magnetic scatterers and $w(k,\phi)$ is the
probability for an electron with wavevector $k$ to be scattered
over an angle $\phi$. In relation to the differential cross
section, we can write $w(k,\phi)=(\hbar k/m)(d\sigma/d\phi)$,
since $\sigma v \Delta t$ is the probability for an electron with
velocity $v$ to interact with a scatterer with cross section
$\sigma$ in a time interval $\Delta t$.

\subsection{Classical result}
\subsubsection{The magnetoresistance}
In Fig.~\ref{fig:clasmagneto.eps}(a), we plot the
magnetoresistance as function of $l_a/R_{a}$ for various
$R_{b}/R_a$, in units of $\rho_0=(n_0/n_e)(\hbar/e^2)$. This is
obtained by inserting the earlier calculated classical
$d\sigma/d\phi$ into $w(k,\phi)$ of Eq.~(\ref{eq:jonassie}). The
magnetoresistance is zero when $l_a/R_a=0$, because for zero
energy, electrons do not move ($v=0$) and consequently do not
experience any scattering. For small values the magnetoresistance
increases linearly up to a certain value after which it decreases
for increasing $l_a/R_a$. This decrease is due to the fact that
for higher energy, the electrons are less deflected because of a
larger cyclotron radius in the magnetic inhomogeneity. We notice
that for increasing $R_b/R_a$ the magnetoresistance has an
overall increase, which can be explained by considering scattering
on the inner and outer profile and how they influence each other.
For larger $R_b/R_a$, the cyclotron radius of the outer part
increases quadratically in $R_b/R_a$ and therefore electrons will
be less deflected for increasing $R_b/R_a$. However, this is
compensated by the fact that the cross section also increases
linearly with $R_b/R_a$, and consequently scattering on the outer
part has more or less the same impact for different $R_b/R_a$.
Therefore, the increase in the resistance is due to the fact that
for larger $R_b/R_a$, the scattering on the inner and the outer
region can be considered as two separate processes, which only
interfere very little. This is not the case for smaller
$R_b/R_a$, where electrons interacting with the outer part are
more likely to interact also with the inner part, which would
deflect the electron in the opposite way, and thus diminishes the
scattering effect produced by the outer part. In short we can
say, that electrons which interact with the magnetic profile, on
average ``feel" a nonzero magnetic field which increases with
increasing $R_b/R_a$, and this results in an increase of the
magneto- and Hall resistance.

The fact that for higher $R_b/R_a$ the scattering problem can be
seen as two separate scattering processes on different vortices
($R_a$,$B_a$) and ($R_b$,$B_b$), is also reflected in the dip in
the magnetoresistance which arises for higher $R_b/R_a$, e.g.
$R_b/R_a=10$. Actually, the magnetoresistance has two peaks of
which one, the lower energy peak, corresponds to scattering on the
outer region, while the second peak corresponds to scattering on
the center region. For higher $R_b/R_a$, the first peak will
shift towards smaller $l_a/R_a$, while the other peak remains in
the same position, and consequently the two scattering processes
will become more distinct.

\subsubsection{The Hall resistance}
The corresponding Hall resistance is plotted in
Fig.~\ref{fig:clasmagneto.eps}(b). We notice that both the
qualitative and quantitative behaviour are more sensitive to the
ratio $R_b/R_a$ than was the case for the magnetoresistance. There
are two striking features: (a) again there is an overall increase
of the Hall resistance with increasing $R_b/R_a$, and (b) for
small $R_b/R_a$, i.e. $R_b/R_a<1.73$, the Hall resistance can
change sign as function of $l_a/R_a$, i.e., as function of the
Fermi-energy or the magnetic field strength.

The fact that the Hall resistance can change sign when
$R_b/R_a<1.73$ is a consequence of the interplay between
scattering in the inner and the outer region of the magnetic
field profile. For infinitesimally small $l_a/R_a$, the outer part
will act as a hard wall and consequently there is no Hall
resistance. For increasing $l_a/R_a$, the electrons will
penetrate deeper in the outer region, but not yet in the inner
core as long as $l_a/R_a$ is small enough, i.e., $2l_b<(R_b-R_a)$
[see the trajectory given by the bold curve in the right figure of
Figs.~\ref{fig:trajectories.eps}(a)]. The Hall resistance is that
due to a vortex with magnetic strength $B_b$. For increasing
$k/B_a$, the electrons will be able to penetrate into the inner
region, where they get deflected to the other side and the Hall
resistance consequently changes sign [see the bold curve in
Figs.~\ref{fig:trajectories.eps}(b,c)].

The reason that this effect does not occur for higher $R_b/R_a$
values, is due to the fact that in these cases, there is only
little interplay between scattering in the inner and the outer
region, as mentioned earlier. It is much more unlikely that an
electron which initially only felt the outer region, for higher
$l_a/R_a$ will interact with the (relatively small) core, and be
swept to the other side. Therefore the Hall resistance for large
$R_b/R_a$ always has the same sign, as generated by the outer
region.

\subsection{Quantum mechanical result}
After insertion of Eq.~(\ref{eq:shittie}) into
Eqs.~(\ref{eq:jonassie}) and (\ref{eq:jonassie2}), we can rewrite
the magnetoresistance and the Hall resistance as function of the
phase shifts
\begin{mathletters}
\begin{eqnarray}
\rho_{xx}=\frac{n_0}{n_e}\frac{\hbar}{e^2}\sum_{m=-\infty}^{\infty}
2\sin^2(\delta_m-\delta_{m+1}), \\
\rho_{xy}=\frac{n_0}{n_e}\frac{\hbar}{e^2}\sum_{m=-\infty}^{\infty}
\sin[2(\delta_m-\delta_{m+1})].
\end{eqnarray}
\end{mathletters}
In Fig.~\ref{fig:res_clas_quan.eps}, we plot the Hall and
magnetoresistance as function of $k/B_a$ for $B_a/B_0=20$ and
$R_b/R_a=1.5$. The solid curve is the quantum mechanical result,
the dashed curve is our previously obtained classical result. We
observe many resonances, which diminish for increasing $k/B_a$.
Except for that, on the average there is rather good agreement
between both curves, which is due to the choice of $B_a/B_0$
being large, and consequently in this figure $k$ --and
consequently the energy-- is large.

There are two types of resonances: (1) those which occur at the
energy of the Landau levels of the inner core of the magnetic
field profile (thin dotted curves), and (2) those corresponding to
quasi-bound snake orbit states.

In case of the first type, the Hall resistance decreases abruptly
(i.e. it has a sawtooth behaviour), while the magnetoresistance
increases. This can be viewed in the right inset in
Fig.~\ref{fig:res_clas_quan.eps}. The reason for this is that at
the energy of the Landau levels (indicated by the vertical dotted
lines), electrons are (quasi-) bound into cyclotron orbits, and
hence cannot (a) contribute to the conduction, and consequently
the magnetoresistance increases, and (b) cannot pile up and
generate a voltage difference on the left and right side, and
consequently the Hall resistance decreases.

For larger energies, the cyclotron orbit in the center will
increase and exceed the inner core classically at $k/B_a=1$.
Nevertheless, quantum mechanically the electron will ``feel" the
presence of the outer magnetic field, and this will change the
resonant energies. The resistance jumps are also less explicit
for higher Landau levels, which is due to the reduced number of
states which fit in the core as a result of the large cyclotron
radius.

Apart from resonant states at the Landau levels, the
magnetoresistance exhibits very sharp peaks, which correspond to
the snake orbit states mentioned earlier. This second type of
resonances can have a lower energy than the first Landau level of
both the inner and outer part, as explained before. Since for all
the snake orbit resonances only one quasi-bound state is involved
(in contrast to the Landau states, where various quasi-bound
states exist for various $m$-values, see
Figs.~\ref{fig:effpot.eps} and \ref{fig:phase.eps}), these peaks
are very sharp (see e.g. the left inset of
Fig.~\ref{fig:res_clas_quan.eps}) and are superimposed on a more
continuous background.

The influence of the quasi-bound snake orbit states is also
visible in the Hall resistance, but not as pronounced as in the
magnetoresistance. These quasi-bound states produce small changes
in the slope of in the Hall resistance $\rho_{xy}$, as is clear
from the dashed curve in the left inset of
Fig.~\ref{fig:res_clas_quan.eps}.

\section{Conclusions}
In this paper we studied scattering on circular symmetric magnetic
field profiles with zero mean magnetic field in the 2DEG. We
considered scattering on a single profile, both classically and
quantum mechanically, and found different types of quasi-bound
states: the Landau-states in the inner core, and different
quasi-bound snake orbit states. Next, we investigated the
diagonal and Hall resistivity in the presence of a randomly
distributed array of these magnetic profiles, using the kinetic
Boltzmann equation, for different magnetic field configurations.
We obtain a nonzero Hall resistance although $\langle B_z
\rangle=0$ and showed that the Hall resistance can change sign as
function of the Fermi-energy or the magnetic field strength. We
found that the electron resonances in the individual magnetic
field profiles are reflected in the Hall and magnetoresistance.

There are two competing effects which make experimental
measurement of these resonances difficult: in order to detect
these quasi-bound states, it is necessary that the energies
(bound or resonant) are not too close to each other. To obtain
this, one has to make a very small magnet (since $E_0=
\hbar^2/mR_a^2$), but then one encounters the problem that in
order to bind the electrons in such a small area one needs a very
strong inhomogeneous magnetic field ($B_0=c\hbar/eR_a^2$), and
currently no magnetic materials are available which can realize
these strong fields.

An example of such a system is the one by Dubonos \emph{et
al.}\cite{dubonos99}: they managed to deposit a single Dy-magnet
with radius $\approx0.1\mu$m on top of a heterojunction,
containing a 2DEG. For this system, our units are given by
$E_0=7.63\times 10^{-3}$ meV and $B_0=0.066$ T. The Fermi energy
in their system was about $E_F=17.86$ meV, which in our units is
$E_F=2341 E_0$ or $k_F\approx 70$. According to their paper, the
stray field could locally generate magnetic fields of $B_a\approx
1T\approx 15 B_0$, which corresponds to $k_F/B_a\approx4.7$, for
which we are in the classical regime and the scattering process
can be calculated classically.

In order to measure individual snake orbits, it is therefore
necessary to include another confinement, p.e. electrical
confinement, which discretizes the energies and makes measurement
possible, as was also used in the paper by Nogaret \emph{et
al.}\cite{nogaret00}

\begin{figure}[t]
\caption{The magnetic field profile (solid curve) and the
theoretical model (dotted curve) for a magnet of radius $R$ and
thickness $d/R=1$, deposited a distance $h/R=0.01$ above a 2DEG,
as shown in the inset. The lower inset shows a simplified top
view of the magnetic field profile in the plane of the 2DEG.}
\label{fig:magnetic.eps}
\end{figure}

\begin{figure}[h]
\caption{This figure corresponds with the situation $B_a/B_0=20$
and $R_b/R_a=1.5$. \emph{On the left:} The differential cross
section for different $k/B_a=l_a/R_a$, i.e., (a) $l_a/R_a=0.5$,
(b) $l_a/R_a=1$, and (c) for $l_a/R_a=2$. The bold curve
corresponds to the classical result, the thin curve to the
quantum mechanical result. \emph{On the right:} Some classical
trajectories interacting with the magnetic profile, resulting in
the differential cross sections on the left. The trajectories
giving rise to the different parts of the different structures,
are grouped schematically.} \label{fig:trajectories.eps}
\end{figure}

\begin{figure}[tb]
\caption{The effective potential $V_{eff}(r)$ as function of
$r/R_0$ for various $m$-values, for the case $B_a/B_0=20$ and
$R_b/R_a=1.5$. The horizontal lines correspond to the resonant
energies.} \label{fig:effpot.eps}
\end{figure}

\begin{figure}[tb]
\caption{The phase shift $\delta_m$ as function of $k/B_a$ for
different $m$-values, for the situation with $B_a/B_a=20$ and
$R_a/R_b=1.5$. Phase jumps of $\pi$ correspond to resonant
states.\vspace{-0.5cm}} \label{fig:phase.eps}
\end{figure}

\begin{figure}[tb]
\caption{The effective potential $V_{eff}(r)$ as function of
$r/R_0$ for $m=-8$ and $m=-21$ when $B_a/B_0=20$ and
$R_a/R_0=1.5$, together with the radial wavefunctions (dotted
curves) at the resonant energies. These quasi-bound states
correspond to different types of snake orbits, propagating
parallel to the magnetic edge, as depicted schematically in the
inset of the figure. } \label{fig:boundstate.eps}
\end{figure}

\begin{figure}[t]
\caption{The cross section $\sigma$ as function of $k/B_a$ in the
classical limit (dashed curve) and if calculated quantum
mechanically (solid curve) for $B_a/B_0=20$ and $R_a/R_0=1.5$.
The marked resonant states correspond to the ones shown in
Fig.~\ref{fig:boundstate.eps} and
Fig.~\ref{fig:res_clas_quan.eps}. Vertical dotted lines are the
Landau energies of the inner core of the magnetic field profile.}
\label{fig:tot_cross.eps}
\end{figure}

\begin{figure}[tb]
\caption{The magnetoresistance (a) and the Hall resistance (b) in
the classical limit as function of $l_a/R_a$ for different
$R_b/R_a$-configurations.} \label{fig:clasmagneto.eps}
\end{figure}

\begin{figure}[tb]
\caption{The magneto- and Hall resistance as function of $k/B_a$
in the classical limit (dashed curves) and if calculated quantum
mechanically (solid curves) for $B_a/B_0=20$ and $R_a/R_0=1.5$.
The marked resonant states correspond to the ones shown in
Fig.~\ref{fig:boundstate.eps} and Fig.~\ref{fig:tot_cross.eps},
and the vertical dotted lines are the Landau energies of the inner
core of the magnetic field profile. The insets show enlargements
of a quasi-bound snake orbit resonance (left) and a Landau level
resonance (right) where the solid curve corresponds to $\rho_{xx}$
and the dashed curve is $\rho_{xy}$.}
\label{fig:res_clas_quan.eps}
\end{figure}

\end{document}